\documentclass[letterpaper]{article} %
\usepackage{aaai25}  %
\usepackage{times}  %
\usepackage{helvet}  %
\usepackage{courier}  %
\usepackage[hyphens]{url}  %
\usepackage{graphicx} %
\usepackage{natbib}  %
\usepackage{caption} %
\usepackage{algorithm}
\usepackage{tikz,graphicx,xcolor}
\usepackage{gensymb}
\usepackage{textcomp}
\usepackage{float,dblfloatfix}
\usepackage{verbatim,array,enumitem}
\usepackage{algorithm,algpseudocode,breqn,amsmath}
\usepackage{adjustbox,diagbox,threeparttable,tablefootnote,colortbl}
\usepackage{booktabs,tabularx,multirow,multicol,hhline}
\usepackage{caption,subcaption}
\PassOptionsToPackage{hyphens}{url} 
\usepackage[capitalise,noabbrev]{cleveref} 
\crefname{appsec}{Appendix}{Appendices} 
\usepackage[misc]{ifsym}
\usepackage[utf8]{inputenc}
\usepackage{booktabs}
\usepackage{lipsum}
\usepackage{tabularray}
\usepackage{longtable}
\usepackage{mdframed}
\usepackage{tcolorbox}
\usepackage{multicol}
\usepackage{bibentry}

\usepackage{amssymb}
\usepackage{svg}
\usepackage{listings}
\usepackage{ragged2e} 

\usepackage{newfloat}
\usepackage{listings}
\DeclareCaptionStyle{ruled}{labelfont=normalfont,labelsep=colon,strut=off} %
\floatstyle{ruled}
\newfloat{listing}{tb}{lst}{}
\floatname{listing}{Listing}
\newcommand{\name}{\textit{MagnetDB}}

\usepackage{xcolor}
\newcommand{\answerYes}[1]{\textcolor{blue}{#1}} 
 
\newcommand{\answerNA}[1]{\textcolor{gray}{#1}}

\title{MagnetDB: A Longitudinal Torrent Discovery Dataset\\with IMDb-Matched Movies and TV Shows}

\begin{document}

\author{
    Scott Seidenberger, Noah Pursell, Anindya Maiti
}
\affiliations{
    University of Oklahoma \\
    seidenberger@ou.edu, noah.a.pursell-1@ou.edu, am@ou.edu
}

\maketitle

\begin{abstract}

BitTorrent remains a prominent channel for illicit distribution of copyrighted material, yet the supply side of such content remains understudied. We introduce \name, a longitudinal dataset of torrents discovered through the BitTorrent DHT between 2018 and 2024, containing more than 28.6 million torrents and metadata of more than 950 million files. While our primary focus is on enabling research based on supply of pirated movies and TV shows, the dataset also encompasses other legitimate and illegitimate torrents. By applying IMDb-matching and annotation to movie and TV show torrents, \name~facilitates detailed analyses of pirated content evolution in the BitTorrent network. Researchers can leverage \name~to examine distribution trends, subcultural practices, and the gift economy within piracy ecosystems. Through its scale and temporal scope, \name~presents a unique opportunity for investigating the broader dynamics of BitTorrent and advancing empirical knowledge on digital piracy. %

\end{abstract}

\section{Introduction}
\label{sec:introduction}

Digital piracy has long posed a challenge to content publishers, policymakers, and internet service providers around the globe~\cite{uschamber_digital_piracy_impact,danaher2020piracy}. While digital piracy can manifest through numerous channels, BitTorrent~\cite{bittorrent,pouwelse2005bittorrent} stands out as one of the most prevalent networks used to facilitate unauthorized peer-to-peer (P2P) sharing of music, films, software, and other media~\cite{digitalmusicnews_verizon_copyright_lawsuit,lifewire_streaming_piracy,cybernews_torrenting_us_adults}. Movies and television shows in particular dominate a significant portion of the pirated content~\cite{wang2010measurement}, reflecting the ongoing supply and demand for entertainment multimedia. Streaming service fragmentation, where exclusive content is spread across multiple platforms, has recently led to increased piracy as consumers become frustrated with managing and affording numerous subscriptions~\cite{ellis2023streaming,bode2018rise,lifewire_streaming_prices_piracy}. BitTorrent's decentralized architecture, ease of use, and recent innovations like streaming capabilities through WebTorrent~\cite{webtorrent_website} sustain its role as a prominent platform for unauthorized distribution, creating a fascinating sociotechnical ecosystem for studying its operation and evolution over time.

In this work, we introduce \name, a large torrent database continuously capturing long-term, real-world BitTorrent activity with explicit focus on the supply side of pirated media. Our motivation aligns with the broader social media and web research communities, which can benefit from examining the cultural and social dynamics behind media consumption, distribution, and public engagement with unauthorized content. The potential applications of \name~are numerous. Cultural analytics researchers can study the availability and popularity of movies or TV shows and how these patterns relate to production timelines or cultural events. Linguists and anthropologists may look at naming conventions and tagging practices within torrent metadata, examining how language usage reflects subcultural identities across different release groups and communities. For policymakers and industry practitioners, the dataset can inform anti-piracy strategies, guide legal service offerings, or provide a more accurate picture of the scope and scale of unlicensed distribution.
With a focus on the supply side, \name~also enables unique investigations into the motivations and behaviors of torrent creators (encoders), shedding light on ``gift economy'' dynamics within piracy ecosystems~\cite{huizing2014explaining}.

\name~spans more than five years of torrent discovery (December 2018 – September 2024), containing \textbf{28,606,694 torrents} with \textbf{950,660,089 files}, which total \textbf{82.87 petabytes} of shared media. 
Although \name~contains torrents across the broader BitTorrent ecosystem, our primary focus is to enable research based on the supply of pirated movies and TV shows. To achieve this, we additionally employ a metadata matching process and identified \textbf{1,562,573} movie and TV show video files covering \textbf{78,740} unique titles on IMDb~\cite{imdb}, and annotated those files with corresponding IMDb identifiers within \name. By cross-referencing pirated video files with rich IMDb metadata, such as genre, release year, ratings, cast, runtime, and reviews, \name~enables more granular analyses of distribution patterns, popularity trajectories, and usage behaviors that surpass what can be inferred from file-level information alone. 

While a significant portion of BitTorrent usage involves unauthorized file sharing, it is also important to recognize that many torrents facilitate legitimate uses, including open-source software releases, academic datasets, and other legal media distributions. Nonetheless, by encompassing both pirated and legitimate torrents, \name~enables a more comprehensive exploration of file-sharing practices, usage trends, and broader distribution dynamics in the BitTorrent network.

\section{The Scene}

Often referred to as ``The Scene,'' this loosely knit yet highly organized network of groups and individuals has played a pivotal role in the distribution of pirated content since the days of dial-up bulletin board systems. Historically, The Scene has been most closely identified with the so-called ``warez'' community, which initially focused on cracking and distributing software. Over time, their activities expanded to music, films, television episodes, e-books, and other digital media. Although The Scene’s structure can appear disjointed from the outside, most participants adhere to an implicit code of conduct and a federated organizational structure designed to promote a sense of community among its affiliates and produce quality warez \cite{goldman2005challenges, warezwars}.

\textbf{Key Players and Motivations.} Scene groups typically occupy specialized roles in the supply chain: ``release groups'' focus on sourcing and preparing new content (e.g., by obtaining pre-release copies of software or media), while ``couriers'' and other intermediaries distribute the files across private platforms, racing to achieve ``zero-day'' status, making content available on or before the official commercial release. In this environment, prestige is closely tied to speed: the faster a group can acquire and release a highly anticipated title, the more its reputation grows within the community \cite{goldman2005challenges}. 

As \name~provides comprehensive data on the torrent supply, it can provide significant insight into the output of the release groups. It is important to note some key taxonomic definitions, as there is an important distinction within the release groups. The groups or individuals that \textit{create} the torrent are the ``encoders'', and the groups that \textit{distribute} the torrent information are the ``sites'' (websites). \name~makes a distinction between these two entities in the dataset.

This subculture operates largely as a gift economy, where members thrive on recognition and status rather than monetary gain \cite{hetu2012welcome, rehn2004politics}. For many participants, the act of collecting and sharing files is itself the reward, having a full library of releases is a point of pride, irrespective of whether those files are ever personally used or consumed.

\textbf{Cultural Underpinnings: Gift Economy and Reputation.} Unlike purely profit-driven piracy rings, The Scene is upheld by a tradition of reciprocity and social capital accumulation. Members who consistently supply high-quality or hard-to-find releases gain respect and clout, which can lead to better access to private servers, exclusive pre-releases, and membership in elite groups. This gift economy ethos is driven by a combination of altruistic sharing norms, social bonding, and competitive one-upmanship. In essence, Scene participants work to cultivate an identity not just as savvy digital sharers, but also as gatekeepers who uphold community standards—particularly regarding file quality, completeness, and timeliness. Notably, Huizing and van der Wal observe that these cultural imperatives can wax or wane over time, as once-lively sub-communities (e.g., the MP3 Scene) either mature, splinter, or face changing technological and social pressures \cite{huizing2014explaining,danaher2020piracy}.

\textbf{Opposition and Enforcement Challenges.} On the other side of The Scene’s shadow economy stand content owners, such film studios, record labels, and gaming companies. They have formed industry consortia who have marshalled significant legal and technological forces to clamp down on unauthorized distribution. Organizations such as the Motion Picture Association (MPA) and the Recording Industry Association of America (RIAA), as well as various government and law enforcement agencies, actively track and pursue major release groups. High-profile investigations have led to global raids, arrests, and lawsuits, but these crackdowns often disrupt only portions of The Scene’s vast and decentralized network. Owing to its resilience and the strong incentives for members to maintain secrecy, new sites, servers, and courier collectives frequently emerge even as others are shut down \cite{basamanowicz2011overcoming}.  

\textbf{Contemporary Developments and Ongoing Tensions.} Despite the persistent threat of legal action, The Scene continues to exert a formative influence on wider piracy practices, including public torrent trackers. Advances in encryption, decentralized storage, and VPN services have enabled the continuation of these piracy practices despite ever-intensifying enforcement. Newer decentralized P2P networks, such as IPFS, are being used to host and distribute pirated content \cite{shi2024closer}.

As David \cite{david2017sharing} argues, free-sharing networks operate at near zero-marginal cost, thereby subverting the core scarcity principle of market-based economies and perpetuating what he terms ``a crime against capitalism.'' This tension between decentralized file-sharing, premised on unconstrained replication, and industry-backed intellectual property regimes underscores the ongoing challenges of enforcing exclusivity in a post-scarcity environment.

The Scene exemplifies the interplay between technological innovation, cultural norms, and external pressures that shape file-sharing ecosystems. Its participants cultivate a reputation-based hierarchy that rewards efficiency, while those seeking to control or criminalize their activities face the practical challenge of shutting down a network whose primary currency is prestige rather than profit. As such, The Scene’s influence remains visible not only in the rapid release of popular media but also in the evolving legal, technical, and social landscapes surrounding file sharing more broadly. 

\section{Background and Related Work}
\label{sec:related}

\textbf{BitTorrent and the DHT.}
A basic unit of distribution on BitTorrent is the \emph{torrent}, an informational file or magnet link containing metadata such as filenames, sizes, and cryptographic hashes. Tying together \emph{swarms} of peers, each torrent spreads through a cooperative process wherein \emph{seeders} (users possessing the complete file) and \emph{leechers} (users still downloading) exchange pieces of the shared file. Although many trackers remain popular for indexing torrents~\cite{piratebay,1337x,rarbg,limetorrents}, including historic torrents and torrents without any network activity, they often have limited coverage of real-world torrent activity as they depend on the encoder to add trackers to their torrents. To address this limitation and obtain a real-world view of actively shared content, we utilize the BitTorrent Distributed Hash Table (DHT) to discover new torrents directly from the P2P network, ensuring that our dataset captures a broader and more dynamic range of content as it emerges. \emph{While BitTorrent DHT enhances resiliency and scale, it also complicates attempts to study the network comprehensively, requiring torrent discovery from a diverse set of peers (which utilizes significant bandwidth and compute) and over extended periods.} BTDigg~\cite{btdigg}, a search engine for torrents discovered on the DHT, provides a useful means to explore decentralized torrent activity, though its dataset is not readily available for direct analysis.

\textbf{Analysis of Peer Activity and Torrents.} 
Despite prior research on piracy in the BitTorrent network, our understanding of the \emph{supply side} of pirated content remains limited. Existing studies focus on tracking seeders and leechers~\cite{zhang2011unraveling,le2010spying} or on the DHT infrastructure as it tracks P2P swarm behaviors~\cite{wang2013measuring}. However, these works often lack comprehensive content metadata and do not fully capture the complexities of what exactly is being shared and by whom. Prior works that do analyze content metadata~\cite{wolchok2010crawling} typically fail to provide a longitudinal scope for observing shifts in piracy trends over several months and years. Consequently, there is a need for a large-scale dataset integrating long-term observation with rich content metadata in order to advance the empirical understanding of piracy factors in the BitTorrent network.

\textbf{Datasets with Torrent Metadata.} 
Only a few existing datasets provide a long-term, consistent view of torrent discovery. The Torrent Metadata Archive hosted on Archive.org~\cite{archive_torrent_metadata} contains metadata for 83 million torrents discovered between 2016 and 2023, though details about its collection process and consistency are unclear. The Kiwi Torrent Research dataset~\cite{kiwi_torrent_research,kiwi_torrent_research_academic_torrents} combines torrents from three different sources, including the Archive.org dataset and an earlier checkpoint of our \name~dataset, highlighting \name~as a recognized and significant contribution to the research community. %
As comprehensive and well-documented torrent datasets remain scarce, \name~stands out as an ongoing effort emphasizing consistency in torrent discovery.

\section{Dataset and Collection Methodology}
\label{sec:dataset}

\subsection{Torrent Data}

To discover torrents and record their metadata, we deployed the open-source BitTorrent DHT crawler \texttt{magnetico} on our self-hosted vantage point~\cite{alper2020magnetico}. This software continuously traverses the decentralized BitTorrent DHT, iteratively seeking out nodes in the network and aggregating torrents discovered from each node. As new nodes are learned, the crawler systematically explores their advertised torrents, building a comprehensive index of accessible swarm information. To enhance the crawler's reach and ensure a broader exploration of the DHT, our \texttt{magnetico} instance is configured with a customizable parameter \textit{indexer-max-neighbors=10000}, enabling it to track and index torrents from up to 10,000 peers simultaneously at any given time. 

\begin{figure}[h]
    \centering
    \includegraphics[width=\linewidth]{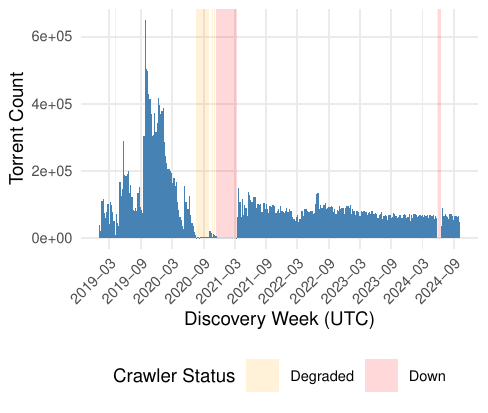}
    \caption{Histogram of discovered torrents with crawler downtime (red) and degraded periods (orange). Degraded is defined as weeks where less than 10,000 torrents were discovered. Uptime over 5 years and 9 months was 93.7\%.}
    \label{fig:discovered}
\end{figure}

Over the 300-week collection period—from December 30, 2018, through September 29, 2024—we gathered \textbf{28.6 million torrents} consisting of more than \textbf{950.6 million individual files}. \Cref{tab:final-counts} summarizes our final collection counts and the subsequent processing steps. Throughout this time, our crawler operated at an \textbf{overall uptime of 93.7\%}, with occasional lapses due to hardware maintenance and network outages. When the system is down, it did not collect new data (shown in red in \cref{fig:discovered}), while degraded periods (shown in orange) reflected reduced coverage, defined here as weeks where fewer than 10,000 new torrents were discovered.

\Cref{fig:discovered} illustrates how the incoming torrent discovery rate initially spiked at the outset of our deployment, reflecting the so-called ``burn-in'' period. During burn-in, \texttt{magnetico} rapidly indexes pre-existing torrents advertised by the DHT. Afterward, discovery stabilizes around a more consistent, long-term rate, capturing newly published torrents as they appear. Research on BitTorrent suggests that the network behaves closest to a random graph, rather than a scale-free or small-world network \cite{su2013measurement}. However, the network still exhibits a small diameter ($< 6$), which helps in the propagation of new information especially on the DHT. Thus, while we consistently observe that most new torrents appear in our database shortly after their initial release, some swarms remain difficult to reach if they have relatively few or transient peers. Manual verification of \name~against known publication times on major torrent distribution sites (such as YTS) confirms that new torrents typically appear in our database on the same day they are released.

Notably, operating a resource-intensive system of this scale demands substantial internet bandwidth, approximately \textbf{30 TB per month}, with a roughly 2:1 ratio of downloaded to uploaded data. While this ensures deeper coverage of the DHT, researchers aiming to replicate or extend this work should carefully consider both the computational cost and the network-level impacts of large-scale crawling.

\subsection{File Processing}

Each torrent can have several files associated with them. In the case of multimedia, the files can be a combination of media files and non-media files. Non-media files can include a text-based ``credit'' file with the information of the uploader, subtitles, or metadata files about the media files. For TV shows, there may be several episodes each as a file in a single torrent. For movies, the movie is typically a single file in the torrent, but in some cases there may be different versions of the same movie in the same torrent, or a sample of the movie as a separate file. Since we focus on multimedia in the form of movies and TV shows, we first identify whether each file is a video file by its file extension. %

As discussed earlier in our section on \textit{The Scene}, suppliers of the bulk of multimedia torrents are known to be relatively well-organized groups, collectively affiliated with The Scene subculture. This subculture voluntarily abides by strict conventions in the creation and naming of pirated content \cite{basamanowicz2011overcoming, dementiev2016bayesian}. A democratic, small council of the top suppliers of pirated content regularly publish the accepted standards that groups follow, and we find consistency with these standards in our dataset. We use these standards to parse the torrent files for the title as well as all the metadata fields.

\newtcolorbox{boxH}{
    sharpish corners,
    boxrule=0pt,
    leftrule=4.5pt
}

\begin{boxH}
    \footnotesize
    \textbf{Movie}\\
    \texttt{Feature.Title.<YEAR>.<TAGS>.[LANGUAGE]} \\
    \texttt{.<RESOLUTION>.<FORMAT>-GROUP}\\
    \vspace{0.5em}

    \textbf{TV Show}\\
    {\scriptsize
    \texttt{Weekly.TV.Show.[COUNTRY\_CODE].[YEAR]}\\
    \texttt{.SXXEXX[Episode.Part].[Episode.Title]}\\
    \texttt{.<TAGS>.[LANGUAGE].<RESOLUTION>.<FORMAT>-GROUP}
    }

    \vspace{1em}
    \hrule
    \vspace{0.3em}
    \centering
    \url{https://scenerules.org/}
\end{boxH}

Using a library that leverages these patterns \cite{parse-torrent-name}, we attempt to parse from each torrent file's name as many metadata fields as possible. While \cref{tab:categorization} shows examples of the metadata fields extracted, the complete list can be found in the documentation \texttt{README} with \name. 

\begin{table}[h!]
\centering
\caption{Example of types of metadata extracted from file names and torrents.}
\begin{tabular}{|l|l|l|}
\hline
\multicolumn{1}{|c|}{\textbf{Content}} 
  & \multicolumn{1}{c|}{\textbf{Distribution}} 
  & \multicolumn{1}{c|}{\textbf{Technical}} \\ \hline

documentary        & site       & audio        \\
3d                 & website    & codec        \\
genre              & network    & resolution   \\
language           & repack     & widescreen   \\
directorsCut       & readnfo    & hdr          \\
internationalCut   & internal   & fps          \\
unrated            & region     & bitDepth     \\
remastered         & encoder    & fps          \\

\hline
\end{tabular}
\label{tab:categorization}
\end{table}

\subsection{Title Matching and Metadata Extraction}

\begin{figure*}[h]
    \centering
    \includegraphics[width=\textwidth]{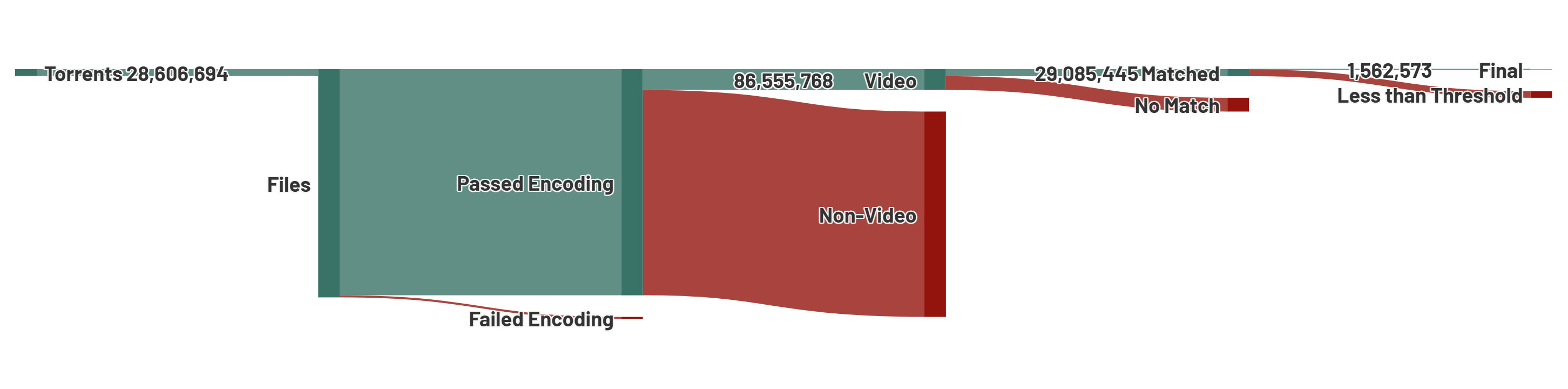}
    \caption{Sankey diagram of the data processing pipeline going from torrents to the final matched dataset. \name~provides all data at each processing step so that other researchers can tailor their own data pipeline to fit their use case.}
    \label{fig:sankey}
\end{figure*}

The next step in the analysis of these multimedia files is to match them against the IMDb dataset \cite{imdb}. Accurate matching allows us to verify the release dates of the content and narrow-down our area of interest to movies and TV shows. We populated an Elasticsearch index with the IMDb dataset. The index was configured with a custom analyzer designed to optimize the matching process for our specific use case.

The custom analyzer employs an edge n-gram filter, which generates n-grams from the beginning of a word up to a specified length. Specifically, we set the minimum n-gram length to 4 and the maximum to 15. This configuration captures meaningful fragments of words while avoiding overly short or excessively long n-grams that might lead to false positives or negatives. The analyzer also incorporates standard text processing steps such as lowercasing and ASCII folding, normalizing the text by converting it to lowercase and replacing accented characters with their ASCII equivalents. This normalization ensures that variations in capitalization or the presence of special characters do not hinder the matching process.  

For each title, the we construct a search query to retrieve the most similar entries from the IMDb index. Elasticsearch's default scoring algorithm, BM25 (Best Match 25) \cite{robertson2009probabilistic}, is used to rank the search results based on their relevance to the query. The BM25 algorithm calculates a relevance score by considering factors such as term frequency, inverse document frequency, and document length normalization. This scoring method effectively ranks potential matches by their textual similarity to the query, which is essential for accurately matching titles that may have slight variations or idiosyncrasies. BM25 scores are positive numbers that can vary widely depending on the query and the dataset and are unbounded. A higher BM25 score indicates a higher relevance between the query and the document. While the choice of threshold is subjective based on the objective of the analysis, the results of the specific implementation of BM25 does not practically impact search effectiveness, with a large scale reproducibility study confirming that BM25 implementations do not yield significant differences between them \cite{kamphuis2020bm25}. 

After retrieving the top matches from the Elasticsearch index, we evaluate the relevance scores to determine if a suitable match has been found. If the highest-ranked IMDb entry has a score exceeding a predefined threshold, we consider it a match and associate the corresponding IMDb metadata with the torrent file. This metadata includes the official title, release year, genre, and other details that enrich our dataset. For the purpose of our initial exploration, we chose a $2\sigma$ threshold match score of 138, illustrated in \cref{fig:match-score}, retaining 1.81\% of the video files, reducing the likelihood of false positives—cases where a torrent title is incorrectly matched to an unrelated IMDb entry due to generic or common terms. For an example of matches with high and low scores, see the example in the Appendix. The entire processing pipeline is illustrated in \cref{fig:sankey} as a Sankey diagram. It is important to note that this data pipeline can be modified to fit any use case, as we provide all the original data.

\section{Descriptive Statistics}
\label{sec:results}

In this section, we provide a broad overview of \name, detailing the scope and composition of the dataset, the characteristics of the files and torrents it contains, as well as the subset of videos successfully matched to external metadata resources like IMDb. \Cref{tab:final-counts} summarizes our key statistics, and the following subsections expand on these figures and the accompanying visualizations. 

\begin{table}[!htb]   %
    \centering
    \caption{Final counts and statistics.}
    \label{tab:final-counts}
    \resizebox{\columnwidth}{!}{%
    \begin{tabular}{l r}
        \toprule
        \textbf{Statistic} & \textbf{Count} \\
        \midrule
        \textbf{\# of torrents} & \textbf{28{,}606{,}694} \\
        \textbf{\# of total files (passed encoding)} & \textbf{942{,}076{,}233} \\
        \# of files with encoding failures & 8{,}583{,}857 \\
        \# of video files & 86{,}555{,}768 \\
        \# of non-video files & 855{,}520{,}465 \\
        \# of matched files (IMDb) & 29{,}085{,}445 \\
        \quad \textit{of which,} \textbf{final (above threshold)} & \textbf{1{,}562{,}573} \\
        \# of matched but below threshold & 27{,}522{,}872 \\
        \midrule
        Match rate (\% of video files matched) & 33.6 \\
        Final match rate (\% of video files, above threshold) & 1.81 \\
        \midrule
        \# of movies & $\approx$ 751{,}256 \\
        \# of TV episodes & $\approx$ 811{,}316 \\
        \midrule
        Avg.\ \# of files per video torrent & $\approx$ 3.026 \\
        \bottomrule
    \end{tabular}%
    }
\end{table}
\subsection{Torrents (Swarms)}

Each torrent, or \emph{swarm}, represents a distinct set of peers simultaneously sharing or downloading a collection of files. Our collection spans \textbf{28.6 million} torrents, containing over \textbf{942 million} individual files which passed encoding checks. \textbf{86.56 million} ($\sim$9.2\%) are identified as video files based on their extensions. As shown in \cref{tab:final-counts}, approximately 3.0\% of all files were discarded during processing due to encoding failures. Across all torrents, we observe a mean of 33.23 files per torrent, illustrating the multi-file structure commonly employed—particularly for large multimedia releases that bundle multiple episodes or file versions.

Should one attempt to download every torrent in \name, the combined size would exceed \textbf{82.87 PB}. Though the vast majority of torrents include only tens or hundreds of megabytes of data, a few contain massive aggregates of multi-terabyte content, such as archival collections.

\subsection{Files}

\begin{figure}[h]
    \centering
    \includegraphics[width=0.95\linewidth]{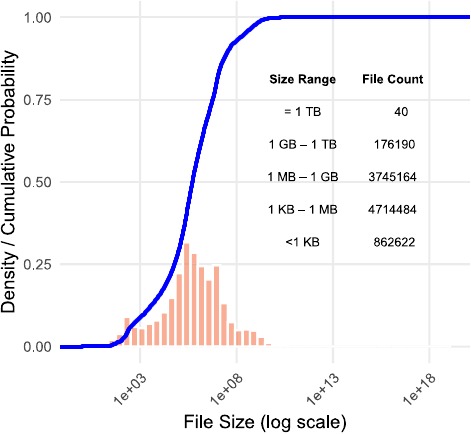}
    \caption{Density plot with overlaid CDF of file sizes in bytes.}
    \label{fig:file-size}
\end{figure}

\Cref{fig:file-size} shows the distribution of the file sizes on a log scale. The histogram portion depicts the density, whereas the line represents the cumulative distribution function (CDF). Most files fall between 1 KB and 1 GB, reflecting typical document, audio, and video content. A non-negligible fraction (over 170,000 files) exceeds 1 GB, attesting to the prevalence of high-definition media content—especially movies and TV episodes, which can range from a few hundred megabytes to tens of gigabytes each. An extreme upper tail (40 files in the range of 1 TB or more) largely corresponds to large archival releases or entire libraries distributed in a single torrent.

\subsection{Videos Matched to IMDb}

\begin{figure}[h]
    \centering
    \includegraphics[width=0.8\linewidth]{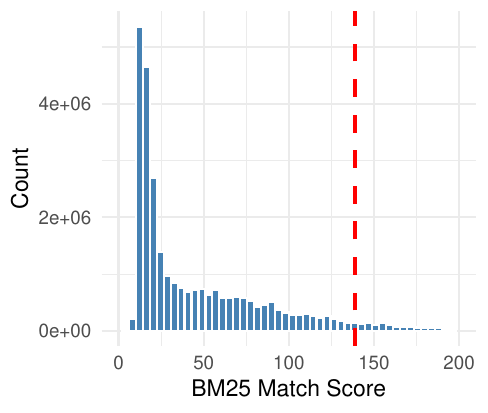}
    \caption{Distribution of BM25 Match Scores to the IMDb Non-Commercial Database. The Red-dashed Line indicates the $2\sigma$ (138) threshold past which we retain rows for the IMDb-matched subset.}
    \label{fig:match-score} 
\end{figure}

Of the 86.56 million video files, around \textbf{29.09 million} ($\sim$33.6\%) were potential matches to IMDb based on the title matching procedure described in the \textit{Title Matching and Metadata Extraction} section. However, to ensure sufficient confidence and minimize false positives, as illustrated in \cref{fig:match-score}, we impose a $2\sigma$ BM25 score threshold of 138 (the red dashed line). This yields a final matched set of \textbf{1.56 million} video files, or $\sim$1.8\% of all videos. These files collectively amount to 1.85 PB of data--just over 2\% of the dataset's total size. Within the final matched dataset, we estimate 751K distinct movies and 811K TV episodes.

\begin{figure}[h]
    \centering
    \includegraphics[width=\linewidth]{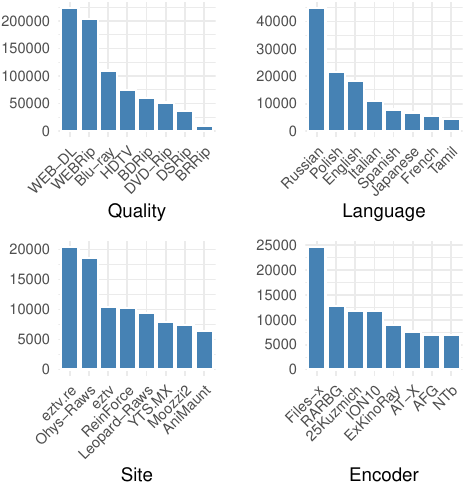}
    \caption{Select breakdowns of the matched video files: Quality labels, language tags, distribution ``sites,'' and encoders (creators of the torrents themselves).}
    \label{fig:breakdowns}
\end{figure}

\Cref{fig:breakdowns} illustrates breakdowns of some of the metadata fields in the matched dataset by video \emph{quality} (e.g., WEB-DL, BluRay), \emph{language} (e.g., English, Russian, Polish), \emph{site} (the groups/websites that distribute the torrent files), and \emph{encoder} (the group or individual responsible for preparing the media). While English and Russian stand out among labeled languages, there is a long tail of other languages. Similarly, a few large release sites and encoders dominate, but countless small or independent groups also contribute to the BitTorrent landscape.

\begin{figure}[h]
    \centering
    \includegraphics[width=1.05\linewidth]{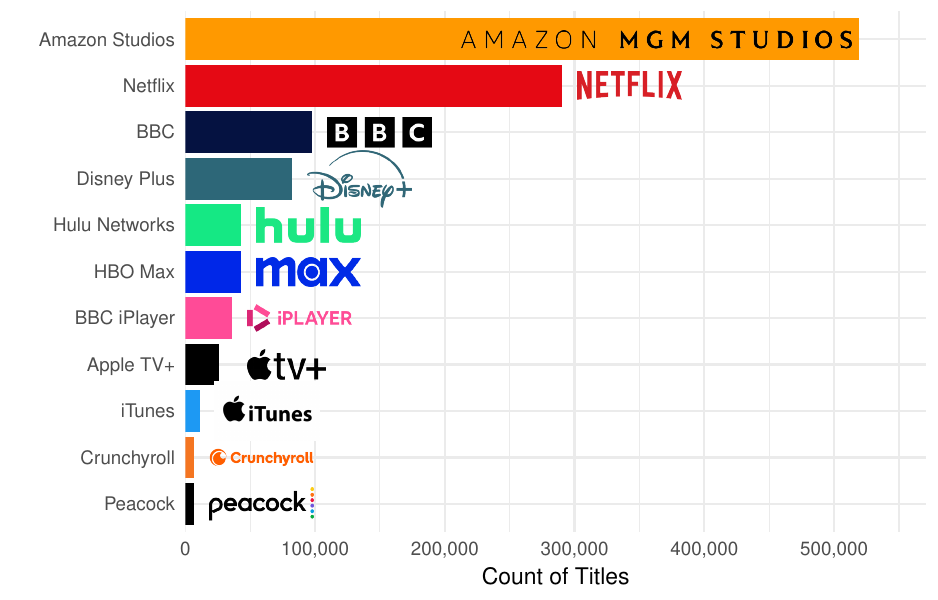}
    \caption{Distribution of torrent-matched titles by streaming service or production network. This data suggests that exclusive or region-locked content fosters higher piracy rates. All logos shown are trademarks of their respective owners.}
    \label{fig:networks}
\end{figure}

\textbf{Targeting of Streaming Services.} \Cref{fig:networks} shows the breakdown of identified titles in \name~by their affiliated streaming service or production company. Notably, large commercial platforms dominate the high end of the spectrum, with \emph{Amazon MGM Studios} leading, followed by \emph{Netflix} and the \emph{BBC}. \emph{Disney Plus}, \emph{Hulu}, and \emph{HBO Max} also occupy substantial niches, illustrating that a range of well-funded, global streaming services face high targeting rates within the BitTorrent ecosystem. These findings underscore the continuing pull of premium digital media in the piracy community, wherein the most prominent and prolific services are also the most widely pirated.

From a supply-side perspective, this aligns with the general pattern that release groups and encoders concentrate on popular or high-demand works. Commercial platforms like Amazon and Netflix invest heavily in exclusive content—often releasing entire seasons of shows at once—which encourages immediate torrenting. Meanwhile, the BBC’s position in the top tier attests to worldwide interest in British-produced series and documentaries, some of which remain difficult to access legally outside the United Kingdom. In essence, these results hint at the interplay between streaming exclusivity, global distribution gaps, and user demand. By examining which platforms rank highest in pirated titles, \name~offers valuable insights into how audience preferences and licensing strategies translate into increased torrent availability.

\textbf{Coverage Over Time.}

\begin{figure}[h]
    \centering
    \includegraphics[width=\linewidth]{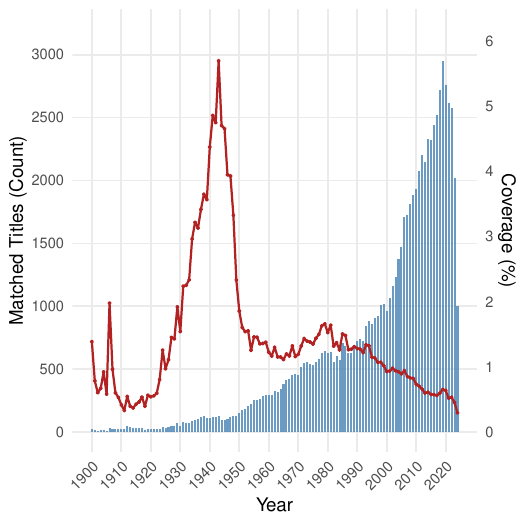}
    \caption{Coverage of \name’s matched subset by release year against IMDb. Blue bars correspond to the absolute count of matched titles per release year; the red line shows percentage coverage relative to IMDb.}
    \label{fig:coverage}
\end{figure}

A central question when evaluating torrent datasets, or any piracy-oriented data, is the degree of coverage in terms of movies and TV shows against a well-known reference, here IMDb. \Cref{fig:coverage} shows how the matched dataset of \name~intersects with IMDb entries over time. The histogram portion indicates the total counts of titles in \name~by release year, and the line shows the percentage coverage relative to IMDB's total corpus for that year. The coverage is below 5\% for most years, showing that there is far more content that has been created than is actively torrented. 

Interestingly, a notable spike occurs in the early-to-mid 1940s. This unexpectedly high coverage in the wartime era likely reflects the enduring popularity of certain classic titles (e.g., \emph{Casablanca (1942)} and \emph{Citizen Kane (1941)}) coupled with the cultural significance of World War II films, which remain actively shared. Additionally, it was in the 1940s that Paramount introduced a new projection system that made color production more affordable. 

For more contemporary releases (post-2000), there is a dramatic surge in IMDb’s catalog, fueled by both increasing global film production and the proliferation of digital-native content such as web series and independent online releases. By contrast, the community of encoders responsible for torrents has not scaled proportionately, and only a fraction of these newer titles appear in \name. This suggests a practical bottleneck on what gets shared via BitTorrent. Scene groups and independent encoders often focus on popular or high-demand works, leaving a large swath of minor or niche titles unrepresented. As IMDb continues to grow--reportedly adding over 10,000 new titles per day--this gap may widen. Future research could employ \name\ to study \emph{why} certain works gain traction in the torrent ecosystem while others remain absent or only appear years after their release, we elaborate on this in our \textit{Discussion}. This gives insight on the torrent network that other datasets and studies do not capture. We hypothesized that \name~would develop large coverage of IMDb, but the data suggests that there are so many more small works that are not captured by our long-standing torrent crawler. It shows that there is so much more to the story on what gets torrented and what gets left out. This dataset challenges preconceived notions about the proliferation of pirated content and offers insights into what content is targeted by torrent suppliers.

\section{Distribution}
\label{sec:distribution}

\name~is FAIR \cite{wilkinson2016fair}: 

\newtcolorbox{boxK}{
    sharpish corners, %
    boxrule = 0pt,
    leftrule = 4.5pt %
}

\begin{boxK}
OSF URL: \url{https://osf.io/9eh47/} \\
DOI: \url{https://doi.org/10.17605/OSF.IO/9EH47} \\

Academic Torrents: \\ \url{https://academictorrents.com/details/1ce8202af7a500469177ed99de5cd9bf66078de0} 
\end{boxK}

\textbf{Findable.} \name~maintains a persistent, globally unique identifier via DOI. The main page on Open Science Framework (OSF) describes the dataset in detail with rich metadata. Furthermore, our data can be shared on other platforms with attribution given its CC BY 4.0 license. 

\textbf{Accessible.} We utilize an OSF data repository for \name~and its successive versions. The protocol and data therein is open and free, and the data is permanently available for public use. For unauthenticated public release, we do not include the magnet links themselves, as \name~contains content identifiers to a significant amount of pirated, copyrighted content. We do not wish to make it easier for a nefarious actor to leverage \name~to find pirated content. However, for academic research purposes we will make the magnet links themselves available upon request to the corresponding author with an authenticated, university affiliated account on OSF. 

\textbf{Interoperable.} The database is delivered as a SQLite database, a stable, cross platform, and one of the most widely deployed database architectures in the world \cite{gaffney2022sqlite}. Additionally, we provide a \textbf{.csv} version of a smaller, curated subset of \name~that focuses specifically on movies and TV shows that we can reasonably match to the IMDB database. This allows \name~to be cross-referenced to other data sources, by torrent, by file, and by other content identifiers such as movie title, for example. 

\textbf{Re-usable.} \name~conforms with our domain-relevant community standards for research on torrents. We support re-usability of the dataset with our choice of license (CC BY 4.0), clear data provenance, and a breadth and depth of data attributes. We have designed and curated this dataset with the purpose of it being used as foundational to future work in this field. Notably, \textbf{updates will be made annually} to the is dataset, with subsequent additions published on the OSF repository. 

\section{Discussion}
\label{sec:discussion}

Our introduction of \name~demonstrates the feasibility and value of a multi-year, large-scale dataset that explicitly focuses on the supply side of BitTorrent. By indexing torrent metadata from a vast swath of the BitTorrent DHT, we have assembled a granular view of what is being uploaded and shared, rather than who downloads or just how swarms evolve. In doing so, our dataset illuminates broader social, cultural, and technical dynamics that underlie piracy ecosystems. Below, we discuss the key insights gleaned from \name, reflect on its ethical implications and limitations, and consider pathways for future research using this resource.

\subsection{Key Insight and Novel Contributions}

\textbf{Longitudinal and Content-Centric View.} Where many existing studies concentrate on peer activity, swarm dynamics, or short-term snapshots of piracy, \name~offers longitudinal perspective spanning over five years. This temporal breadth enables analyses of how supply-side behaviors evolve in parallel with technological, social, and even geopolitical factors. Our integration of file-level metadata with IMDb identifiers gives a richer picture of what exactly is getting uploaded and by whom. Crucially, these linkages enable targeted inquiries into the popularity and availability of specific media types, revealing distribution trends that cannot be captured by focusing solely on tracker-level measurements. 

\textbf{Supply-Side Cultural Dynamics.} By zeroing in on the individuals and groups who create torrents, \name~underscores the influential role of The Scene in shaping what content is made available and how. Our findings support the ida that The Scene's practices, however informal and federated, retain a high degree of consistency and organization. Encoders still label and structure torrents according to well-established naming conventions, a tradition upheld by a subcultural gift economy and its emphasis on reputation, speed, and quality. While The Scene's norms can appear obscure, they significantly impact the broader network. Groups with strong reputations set naming standards that other encoders replicate, demonstrating subcultural capital translates into soft power within this ecosystem. This is a level of analysis only provided by a dataset like \name.

\textbf{Evidence of Selective Coverage.} Although \name~spans tens of millions of torrents, our analysis of matched video files reveals that a substantial fraction of the IMDb catalog is simply never torrented. We suspect this is particularly pronounced for obscure titles with low demand or limited mainstream attention based on what is known about how encoders and sites prioritize the availability of content. Despite ongoing expansions in film and television production, The Scene and associated groups prioritize specific, high-value content, often new releases or classic titles with enduring popularity. Our initial exploration of \name~shows a diminishing proportion of new releases that appear in torrent form over time, which suggests a dynamic interplay between supply-side capacity, user demand, and underlying motivations in this gift economy. This raises intriguing questions about the cultural gatekeeping role performed by these piracy groups, who effectively determine which titles, and in which form, achieve broad accessibility through these P2P networks--exploration of these questions we leave for future work. 

\subsection{Limitations}

\textbf{Potential Bias in Coverage.} The BitTorrent DHT is a constantly changing network, and any given vantage point can miss torrents that either have extremely short lifespans or exist in small, localized subgraphs of the network with low propagation to the rest of the network. While we took a longitudinal and consistent approach, no data collection pipeline can fully index every torrent in existence due to the decentralized topology of the P2P network. Our coverage is therefore best viewed as a robust, but not provably exhaustive, snapshot of real-world activity at scale.

\textbf{Matching Imperfections.} Torrent file names can, and often are, incomplete, misleading, or unhelpfully generic, which complicates any automated matching to external databases such as IMDb. Despite internal validation and conservative BM25 thresholds, spurious matches and missed titles inevitably persist. Additionally, due to how the BM25 algorithm calculates match scores, where the max possible score is positively correlated to the length of text, there is a potential for match bias towards entries who's titles are longer. Researchers using \name~should remain aware of the potential for false positives or negatives and choose their own matching thresholds and conduct data validation checks appropriate for their specific use cases.

\textbf{Ethical and Legal Implications.} While it is well-known that BitTorrent hosts a wide array of copyrighted content, the platform is fundamentally a dual-use technology. It serves many legitimate purposes, such as disseminating open-source software or bypassing censorship in oppressive regimes. Our role in providing \name~is to offer a dataset, that will become an updating catalog of datasets, that can advance empirical research on sociotechnical systems, not to endorse or facilitate copyright infringement or other crimes. We do not distribute magnet links themselves in the publicly accessible version of \name, balancing transparency for academic inquiry with a commitment to minimize unethical or illicit use. 

\subsection{Applications and Future Directions}

\textbf{Cultural Analytics.} The structured metadata in \name~supports diverse lines of inquiry into sociocultural trends. For instance, researchers could examine how films, television shows, and other media travel across linguistic and cultural borders over time. A potential avenue of inquiry involves analyzing which titles are dubbed, subtitled, tagged, or otherwise adapted for different languages. By tracking the emergence of multi-language torrents and the frequency of specific language tags or subtitles, researchers could investigate how certain genres, franchises, or classic works proliferate beyond their original audiences. This form of cultural diffusion reflects both grassroots demand (e.g., fans creating or sharing subtitles) and broader, industry-led strategies to reach new markets.

\name~also facilitates longitudinal studies on the ebb and flow of cultural relevance. Some titles appear for a fleeting moment only to vanish as interest wanes. Others persist in active torrents long after their initial release, sustained by perennial fan enthusiasm, new adaptations, or nostalgic revivals. The dataset allows for comparative analyses of how, for instance, cult classics gradually gain renewed popularity or how once-obscure films become ``resurrected'' due to external triggers (e.g., anniversaries, streaming re-releases, or cultural rediscovery). By correlating torrent availability with societal events and production milestones, scholars can trace the lifecycle of media artifacts, identifying when and why certain works achieve evergreen status while others fade away or abruptly re-emerge. 

\textbf{Policy and Anti-Piracy Strategies.} Given its multi-year breadth, \name~can inform the design and evaluation of anti-piracy interventions. Industry groups and policymakers could, for example, assess whether certain enforcement actions or takedown efforts correlate with measurable changes in torrent supply, even from specific individuals, sites, or collectives. Likewise, our data could support modeling how new offering types, such as simultaneous digital releases, shift the supply of pirated content. In contrast to many existing data sources that rely on user-facing distribution sites, \name~enables direct observation of supply and its trajectories.

\textbf{Evolving Supply of Other Media.} Beyond movies and TV shows, the dataset encompasses torrents for a wide variety of media, from e-books and music to software packages. These alternate modalities follow distinct production, distribution, and consumption practices. Software torrents, for instance, raise unique security concerns due to the risk of embedded malware, a fundamentally different threat model from video files. Studies could therefore examine how software piracy evolves over time, comparing it with other forms of digital media in terms of both scale and supply patterns. Similarly, examining e-book and audio content could yield fresh perspectives on cultural diffusion, user demand for niche or academic material, and shifting norms around digital rights management.

Additionally, \name~captures torrents for a considerable volume of adult content—an area of increasing policy scrutiny in certain jurisdictions. Researchers and policymakers alike could leverage torrent-level evidence to explore how legislative or platform-specific bans on adult content affect its distribution in P2P networks. For instance, if newly passed age-verification laws or website blocks reduce access to certain sites, \name~data may reveal whether BitTorrent sees a compensatory spike in adult content sharing. By mapping these shifts, scholars could assess the efficacy and potential unintended consequences of bans or restrictions that attempt to control adult content consumption online.

\textbf{Beyond BitTorrent: Emerging P2P Networks.} Although BitTorrent remains one of the most widely used P2P protocols, newer networks, such as IPFS, are increasingly being used to distribute illicit content \cite{sokoto2024guardians, shi2024closer}. Future work should leverage \name~to cross-reference if these emerging platforms carry the same supply-side incentives and subcultural norms. The introduction of cryptocurrency-based incentives in file-sharing systems revitalizes an old avenue of research, as the traditional gift economy model may once again hybridize to give way to more profit-oriented operations.

\textbf{Extensions and Updates.} We envision \name~as a \textbf{living resource that will be updated annually.} Repeated longitudinal snapshots of torrent data can illuminate shifts in supply-side practices, including the emergence or decline of particular individuals, groups, sites, and naming conventions. Researchers could also further enrich \name~by integrating additional metadata sources to explore how torrent communities produce, share, and discuss content.

\section{Conclusion}
\label{sec:conclusion}

By making \name~openly available for academic use, we seek to catalyze a new wave of research on piracy, subcultural economies, and P2P data exchange. Our holistic approach, spanning torrent discovery, file-level parsing, and IMDb matching, provides a more comprehensive lens on what is being uploaded and who is uploading it. While digital piracy continues to prompt legal and ethical debates, it remains an enduring sociotechnical phenomenon with implications for information access, creative industries, and digital culture. We believe \name~will prove a valuable springboard for those seeking to understand and interpret these complex dynamics through the lens of real-world, empirical data.

\bibliography{references}

\clearpage
\subsection*{Paper Checklist}
\label{sec:checklist}

\begin{enumerate}
\item For most authors...
\begin{enumerate}
    \item  Would answering this research question advance science without violating social contracts, such as violating privacy norms, perpetuating unfair profiling, exacerbating the socio-economic divide, or implying disrespect to societies or cultures?
    \answerYes{Yes. Our database is aggregated from publicly available magnet links and torrent metadata; no personally identifiable information is collected or released. Potentially sensitive elements (e.g., direct links) are withheld from public release to reduce misuse.}
  \item Do your main claims in the abstract and introduction accurately reflect the paper's contributions and scope?
    \answerYes{Yes. We clearly state that we introduce a large-scale torrent dataset and document its composition, methodology, and potential uses.}
   \item Do you clarify how the proposed methodological approach is appropriate for the claims made? 
    \answerYes{Yes. We explain our approach for collecting torrents, parsing file metadata, and matching titles to IMDb.}
   \item Do you clarify what are possible artifacts in the data used, given population-specific distributions?
    \answerYes{Yes. We discuss naming conventions, potential biases in content availability, and how ``the scene'' culture might affect representativeness.}
  \item Did you describe the limitations of your work?
    \answerYes{Yes. We have a “Limitations” section that addresses potential biases, incomplete coverage, and risk of partial false positives in matching.}
  \item Did you discuss any potential negative societal impacts of your work?
    \answerYes{Yes. We acknowledge that torrent data is often associated with pirated content and discuss ethical considerations.}
      \item Did you discuss any potential misuse of your work?
    \answerYes{Yes. We note that while we do not release magnet links publicly, we remain alert to risks of facilitating unauthorized content access.}
    \item Did you describe steps taken to prevent or mitigate potential negative outcomes of the research, such as data and model documentation, data anonymization, responsible release, access control, and the reproducibility of findings?
    \answerYes{Yes. We describe our access-control policy for magnet links, omit direct links in the public release, and provide a documented process for responsible access.}
  \item Have you read the ethics review guidelines and ensured that your paper conforms to them?
    \answerYes{Yes.}
\end{enumerate}

\item Additionally, if your study involves hypotheses testing...
\begin{enumerate}
  \item Did you clearly state the assumptions underlying all theoretical results?
    \answerNA{NA}
  \item Have you provided justifications for all theoretical results?
    \answerNA{NA}
  \item Did you discuss competing hypotheses or theories that might challenge or complement your theoretical results?
    \answerNA{NA}
  \item Have you considered alternative mechanisms or explanations that might account for the same outcomes observed in your study?
    \answerNA{NA}
  \item Did you address potential biases or limitations in your theoretical framework?
    \answerNA{NA}
  \item Have you related your theoretical results to the existing literature in social science?
    \answerNA{NA}
  \item Did you discuss the implications of your theoretical results for policy, practice, or further research in the social science domain?
    \answerNA{NA}
\end{enumerate}

\item Additionally, if you are including theoretical proofs...
\begin{enumerate}
  \item Did you state the full set of assumptions of all theoretical results?
    \answerNA{NA}
	\item Did you include complete proofs of all theoretical results?
    \answerNA{NA}
\end{enumerate}

\item Additionally, if you ran machine learning experiments...
\begin{enumerate}
  \item Did you include the code, data, and instructions needed to reproduce the main experimental results (either in the supplemental material or as a URL)?
    \answerNA{NA}
  \item Did you specify all the training details (e.g., data splits, hyperparameters, how they were chosen)?
    \answerNA{NA}
     \item Did you report error bars (e.g., with respect to the random seed after running experiments multiple times)?
    \answerNA{NA}
	\item Did you include the total amount of compute and the type of resources used (e.g., type of GPUs, internal cluster, or cloud provider)?
    \answerNA{NA}
     \item Do you justify how the proposed evaluation is sufficient and appropriate to the claims made? 
    \answerNA{NA}
     \item Do you discuss what is ``the cost`` of misclassification and fault (in)tolerance?
    \answerNA{NA}
  
\end{enumerate}

\item Additionally, if you are using existing assets (e.g., code, data, models) or curating/releasing new assets, \textbf{without compromising anonymity}...
\begin{enumerate}
  \item If your work uses existing assets, did you cite the creators?
    \answerYes{Yes. We cite IMDb for the metadata, Magnetico for torrent indexing, and other relevant software libraries.}
  \item Did you mention the license of the assets?
    \answerYes{Yes. We clarify IMDb data usage terms and open-source licenses for any libraries.}
  \item Did you include any new assets in the supplemental material or as a URL?
    \answerYes{Yes. We provide data documentation plus a link to the OSF repository and Academic Torrents link that has all the data.}
  \item Did you discuss whether and how consent was obtained from people whose data you're using/curating?
    \answerNA{We use publicly available torrent metadata and do not include PII, so explicit consent is not applicable.}
  \item Did you discuss whether the data you are using/curating contains personally identifiable information or offensive content?
    \answerYes{Yes. We explicitly note there is no PII and disclaim that while some files may have offensive or adult material, the content itself is not distributed.}
\item If you are curating or releasing new datasets, did you discuss how you intend to make your datasets FAIR?
\answerYes{Yes. \textit{Distribution} section details how we follow FAIR principles.}
\item If you are curating or releasing new datasets, did you create a Datasheet for the Dataset? 
\answerYes{Yes. In addition to what is covered in the main body of the paper, we include full dataset documentation in the \texttt{README} on the data repository.}
\end{enumerate}

\item Additionally, if you used crowdsourcing or conducted research with human subjects, \textbf{without compromising anonymity}...
\begin{enumerate}
  \item Did you include the full text of instructions given to participants and screenshots?
    \answerNA{NA}
  \item Did you describe any potential participant risks, with mentions of Institutional Review Board (IRB) approvals?
    \answerNA{No IRB required, as we do not collect data directly from human subjects.}
  \item Did you include the estimated hourly wage paid to participants and the total amount spent on participant compensation?
    \answerNA{NA}
   \item Did you discuss how data is stored, shared, and deidentified?
   \answerNA{NA}
\end{enumerate}

\end{enumerate}

\appendix
\newpage
\section{Appendix}
\label{appendix:whatever}

\subsection{Examples of a Low and High Match Score}

\newtcolorbox{lowmatchbox}[1][]{%
  sharpish corners,
  boxrule = 0pt,
  leftrule = 4.5pt,
  colback = red!5,
  colframe = red!75!black,
  fonttitle = \bfseries,
  title = {Low Match Score},
  #1
}

\newtcolorbox{highmatchbox}[1][]{%
  sharpish corners,
  boxrule = 0pt,
  leftrule = 4.5pt,
  colback = green!5,
  colframe = green!65!black,
  fonttitle = \bfseries,
  title = {High Match Score},
  #1
}

\begin{lowmatchbox}
\textbf{Filename:} \texttt{Riviera.S02E01.WEBRip.x264} \\
\texttt{-ION10.mp4} \\
\textbf{Candidate Match:} \emph{On the Riviera}\\
\textbf{Correct Title:} \emph{Riviera}\\
\textbf{Match Score:} 18.3

A recurring problem arises from short prepositions and articles 
(e.g., “on,” “the”) that mismatch the true title.
\end{lowmatchbox}

\begin{highmatchbox}
\textbf{Filename:} \texttt{shadowhunters.the.mortal.} \\
\texttt{instruments.s03e20.1080p.web.h264} \\
\texttt{-tbs.mkv} \\
\textbf{Matched Title:} \emph{Shadowhunters: The Mortal Instruments}\\
\textbf{Match Score:} 260

This example highlights a near-ideal match, where the full series 
title appears in the torrent name, leading to a high BM25 score.
\end{highmatchbox}

\end{document}